\documentclass[a4paper,final]{easychair}
\usepackage{listings}
\usepackage{ifthen}
\usepackage{cite}

\DeclareSymbolFont{letters}{OML}{cmbboard}{m}{it}
\SetMathAlphabet{\mathcal}{normal}{OMS}{xmdcmsy}{m}{n}
\hypersetup{%
  colorlinks=true,%
  citecolor=blue,%
  urlcolor=blue,%
  linkcolor=blue,%
}

\theoremstyle{definition}
\newtheorem{example}{Example}[section]

\lstdefinestyle{xml}{%
 language=XML,%
 basicstyle=\upshape\ttfamily,%
 flexiblecolumns=false,%
 basewidth={0.5em,0.45em},%
 tabsize=1,%
 xleftmargin=1em,%
 xrightmargin=1em,%
 mathescape=true,%
 showstringspaces=false,%
}

\newcommand\xml[1]{\text{\lstinline[style=xml]^#1^}}
\newcommand\xtag[2][]{%
  \ifthenelse{\equal{#1}{}}{\xml{<#2/>}}{\xml{<#2>}#1\xml{</#2>}}%
}

\newcommand\formatID[1]{\ensuremath{{\sf#1}}}
\newcommand\CeTA{\formatID{C\kern-0.10exe\kern-0.40exT\kern-0.40exA}}
\newcommand\CiME{\formatID{C\textit{i}ME}}
\newcommand\Coccinelle{\formatID{Coccinelle}}
\newcommand\CoLoR{\formatID{CoLoR}}
\newcommand\Coq{\formatID{Coq}}
\newcommand\Haskell{\formatID{Haskell}}
\newcommand\Isabelle{\formatID{Isabelle}}
\newcommand\IsaFoR{\formatID{Isa\kern-0.10exF\kern-0.10exo\kern-0.10exR}}
\newcommand\OCaml{\formatID{OCaml}}
\newcommand\Rainbow{\formatID{Rainbow}}
\newcommand\SML{\formatID{SML}}
\newcommand\TTTT{\formatID{%
 T\kern-0.2em\raisebox{-0.3em}{$\mathsf{T}$}\kern-0.2emT\kern-0.2em%
 \raisebox{-0.3em}{$\mathsf{2}$}%
}}

\newcommand\id[1]{\ensuremath{\langle\textsl{#1\/}\rangle}}
\newcommand\dby{\stackrel{\scriptscriptstyle\textsf{def}}{=}}

\newcommand\DP{\mathsf{DP}}
\newcommand\RR{\mathcal{R}}
\newcommand\PP{\mathcal{P}}
\newcommand\tcap{\mathsf{tcap}}
\newcommand\Var{\mathcal{V}\mathsf{ar}}
\newcommand\Vars{\mathcal{V}}
\newcommand\YES{\ensuremath{\mathsf{YES}}}
\newcommand\NO{\ensuremath{\mathsf{NO}}}

\author{%
 Christian Sternagel \and René Thiemann \and
 Sarah Winkler \and Harald Zankl \and
 \centerline{University of Innsbruck, Austria}
}
\title{\CeTA{} -- A Tool for Certified Termination Analysis\thanks{%
  This project is supported by FWF (Austrian Science Fund) project P18763.}}
\titlerunning{\CeTA{} -- A Tool for Certified Termination Analysis}
\authorrunning{C.~Sternagel et al.}

\begin{document}
\maketitle

\section{Motivation}
Since the first termination competition%
\footnote{\url{http://termination-portal.org/wiki/Termination_Competition}}
in 2004 it is of great interest, whether
a proof---that has been automatically generated by a termination tool---is
indeed correct. The increasing number of termination proving techniques as well
as the increasing complexity of generated proofs (e.g., combinations of several
techniques, exhaustive labelings, tree automata, etc.), make certifying (i.e.,
checking the correctness of) such proofs more and more tedious for humans.
Hence the interest in automated certification of termination proofs. This led to
the general approach of using proof assistants (like \Coq{}~\cite{Coq} and
\Isabelle{}~\cite{Isabelle}) for certification. At the time of this writing, we
are aware of the two combinations \Coccinelle/\CiME{}~\cite{A3Pat,A3PatGraph}
and \CoLoR/\Rainbow{}~\cite{Color}. Here \Coccinelle{} and \CoLoR{} are \Coq{}
libraries, formalizing rewriting theory. Then \CiME{} as well as \Rainbow{} are
used to transform XML proof trees into \Coq{} proofs which heavily rely on those
libraries. Hence if you want to certify a proof you need a termination tool that
produces appropriate XML output, a converter (\CiME{} or \Rainbow{}), a local
\Coq{} installation, and the appropriate library (\Coccinelle{} or \CoLoR{}).

In this paper we present the latest developments for the new combination
\IsaFoR/\CeTA~\cite{CeTA} (version 1.03). Note that the system design has two major differences in
comparison to the two existing ones. Firstly, our library \IsaFoR{}
(\emph{Isabelle Formalization of Rewriting}) is written for the theorem prover
\Isabelle/HOL and not for \Coq. Secondly, and more important, instead of
generating for each proof tree a new proof, using an auxiliary tool, our library
contains several executable check-functions. Here, \emph{executable} means
that it is possible to automatically obtain a functional program (e.g., in
\Haskell), using \Isabelle's code generation facilities~\cite{codegen07}. For
each termination technique that we have implemented in \IsaFoR, we have formally
proven that whenever such a check is accepted, the termination technique is
applied correctly. Hence, we do not need to create an individual \Isabelle{}
proof for each proof tree, but just call the \mbox{check-function} for checking the
whole tree (which does nothing else but calling the separate checks for each
termination technique occurring in the tree). Additionally, our functions
deliver error messages that are using notions of term rewriting (in contrast to
error messages from a proof assistant that are not easily understandable for the
novice). Furthermore, \IsaFoR{} contains a functional parser that accepts XML
proof trees. Since even this parser is written in \Isabelle{}, we can freely
choose for which programming language we want to generate code (\Isabelle{}
currently supports \Haskell, \OCaml, and \SML). At the moment we generate
\Haskell{} code, resulting in our certifier~\CeTA{} (\emph{Certified Termination
Analysis}). However, for a user of \CeTA{} it will make no difference if it was
compiled from \Haskell{} sources or \OCaml{} sources. 

To certify a proof using
\CeTA{}, you just need a \CeTA{} binary plus a termination tool that is able to
print the appropriate XML proof tree.  Moreover, the runtime of certification is
reduced significantly.  Whereas it took the other two approaches more than one
hour to certify all proofs during the last certified termination competition,
\CeTA{} needs about two minutes for all examples, the average time per system
being 0.14 seconds.  Note that \CeTA{} can also be used for modular
certification. Each single application of a termination technique can be
certified by just calling the corresponding \Haskell{} function.  Another
benefit of our system is its robustness. Every proof which uses weaker
techniques than those formalized in \IsaFoR{} is accepted. For example,
termination provers can use the simple graph estimation of \cite{AG00}, as it is
subsumed by our estimation. 


\IsaFoR{}, \CeTA{}, and all details about our experiments are
available at
\CeTA's website.\footnote{\url{http://cl-informatik.uibk.ac.at/software/ceta}}

\section{Supported Techniques}
\label{sec:techniques}
Currently, \CeTA{} features certifying proofs for term rewrite systems (TRSs),
i.e., the initial problem is always, whether a given TRS~$\RR$ is terminating or
not. Hence on the outermost level of a proof we distinguish between termination
and nontermination. 

\paragraph{Termination.}
There are already several techniques for certifying termination proofs. Those
techniques can be categorized as follows:
\begin{enumerate}
  \item A trivial proof (for empty $\RR$).
  \item\label{remove} Removing some rules~$\RR'$ from~$\RR$ such that
    termination of $\RR\setminus\RR'$ implies termination of $\RR$ \cite{RTA04}.    
  \item\label{DPFrame} Switching to the dependency pair (DP) framework 
  by applying the DP transformation, resulting in the initial
  DP problem $(\DP(\RR),\RR)$.
\end{enumerate}
In case \ref{remove}, monotone linear polynomial interpretations over the
naturals are supported. For \ref{DPFrame}, the following processors are available
to prove finiteness of a DP problem~$(\PP,\RR)$:
\begin{description}
\item[Empty $\PP$:] Emptiness of the $\PP$-component of a DP problem implies
that the problem is finite.

\item[Dependency Graph:] We support a dependency graph estimation that is based
on a combination of \cite{GTS05} and \cite{HM05} (using the function
$\tcap$). We call this estimation EDG***. After the estimated graph is computed,
$(\PP,\RR)$ is split into the new DP problems $(\PP_1,\RR)$, \ldots,
$(\PP_n,\RR)$ (one for each strongly connected component of the estimated
graph). Note that our implementation allows a termination tool to use any
weaker estimation than EDG***, i.e., any estimation producing a
graph which contains at least those edges that are present in EDG***.

\item[Reduction Pair:] In the abstract setting of \IsaFoR{}, the notion
of \emph{reduction pair} has been formalized. For concrete proofs there
are the following instances:
\begin{itemize}
  \item Weakly monotone linear polynomial interpretations over the
  natural numbers with negative constants \cite{HM07}. Those can be used to remove
  rules from $\PP$. Here, only the usable rules \cite{GTS05}---w.r.t.~the 
  argument filter that is implicit in the reduction pair---have to be oriented.

  \item Strictly monotone linear polynomial interpretations over the
  natural numbers which can be used to remove rules from both
  $\PP$ and $\RR$ (where $\RR$ can first be reduced to the usable rules).
\end{itemize}
\end{description}

\paragraph{Nontermination.}
For the time being, loops are the only certifiable way of proving
nontermination. If a TRS is not well-formed (i.e.,
$l\in\Vars$ or $\Var(r) \not\subseteq \Var(l)$ for some rule $l\to r$)
the loop is implicit. Otherwise a loop is represented by a
context~$C$, a substitution~$\sigma$, and terms~$t_1$ to~$t_n$
such that $t_1 \to \cdots \to t_n \to C[t_1\sigma]$.

\section{Use it for Your Termination Prover}
\label{sec:integration}
To use \CeTA{} for certifying your own proofs, you need a termination tool that
generates appropriate XML output plus a \CeTA{} binary (if you want to build
\CeTA{} yourself you will still need an \Isabelle{} installation and the
\IsaFoR{} library, as well as a \Haskell{} compiler). Hence, the main work will
be to modify the termination tool in order to generate XML. In the following we
will first give a short overview of the main components that are currently part
of our XML format and then show how to call \CeTA.
\begin{example}
As an example consider the 
structure of a termination proof for $\RR$,
where first a reduction pair has been used to reduce it to the TRS $\RR'$.
Afterwards the dependency pairs of $\RR'$ are computed, resulting in a DP
\pagebreak
problem. Then the proof proceeds by applying a dependency graph estimation.
\begin{lstlisting}[style=xml]
<proof>
  <ruleRemoval>
    <redPair>...</redPair>
    <trs>$\RR'$</trs>
    <dpTrans>
      <dps>$\DP(\RR')$</dps>
      <depGraphProc>...</depGraphProc>
    </dpTrans>
  </ruleRemoval>
</proof>
\end{lstlisting}
\end{example}
The XML format is structured such that on the one hand, new components (like DPs
after the dependency pair transformation or the reduction pair processor) are
explicitly provided by the user, and on the other hand, the user cannot change
components which must not be changed (e.g., the TRS when applying the DP graph
processor). The general structure of proofs is as
follows:\footnote{\url{http://cl-informatik.uibk.ac.at/software/ceta/xml/ceta.xsd}%
\href{http://cl-informatik.uibk.ac.at/software/ceta/xml/ceta.xsd.pdf}{\texttt{[.pdf]}}}
\[\begin{array}{rcl}
\id{proof}          &\dby& \xtag[\id{trsProof}]{proof}
                     \mid  \xtag[\id{trsDisproof}]{proof}\\
\\[-2ex]
\id{trsProof}       &\dby& \xtag[\id{redPair}\id{trs}\id{trsProof}]{ruleRemoval}\\
                    &\mid& \xtag[\id{dps}\id{dpProof}]{dpTrans}\\
                    &\mid& \xtag{rIsEmpty}\\
\\[-2ex]
\id{dpProof}        &\dby& \xtag[\id{component}^*]{depGraphProc}\\
                    &\mid& \xtag[\id{redPair}\id{dps}\id{usableRules}\id{dpProof}]{redPairUrProc}\\
                    &\mid& \xml{<monoRedPairUrProc>}\id{redPair}\id{dps}\id{trs}\id{usableRules}\id{dpProof} \\
                    & & \qquad \xml{</monoRedPairUrProc>}\\
                    &\mid& \xtag{pIsEmpty}\\
\\[-2ex]
\id{redPair}        &\dby& \xtag[\id{interpretation}]{redPair}\\
\\[-2ex]
\id{interpretation} &\dby& \xtag[\id{type}\id{domain}\id{interpret}^*]{interpretation}\\
\\[-2ex]
\id{trsDisproof}   &\dby& \xtag[\id{substitution}\id{context}\id{term}^*]{loop} \\	               &\mid& \xtag{notWellFormed}
\end{array}\]
For \id{ruleRemoval}, the component \id{trs} holds the rules that could only be
weakly oriented by the given \id{redPair}. For \id{dpTrans}, the dependency
pairs are provided by \id{dps}. The list of \id{component}s within the
dependency graph processor denotes all the strongly connected components
(including trivial ones consisting of a single node without a self-edge) in
topological order. For \id{interpretation}s we currently support as \id{type}
only linear polynomials and as \id{domain} only the natural numbers.
(Detailed descriptions of all the other components can be found on \CeTA's
website.)

\CeTA{} is called with two arguments: the first is the problem for
which a proof should be certified and the second is the corresponding proof.
Hence to certify the proof~\texttt{proof.xml} for the
problem~\texttt{problem.xml}, \CeTA{} is called as follows:
\begin{quote}
\texttt{\$ CeTA problem.xml proof.xml}
\end{quote}
The problem and the proof have to be in XML. For the problem the
proposed XTC format---that should soon replace the TPDB format in the termination
competition---is used.

Before applying a check-function to a given proof, the internal data structure
is converted to XML and compared to the input string. A proof is only accepted
if both are equal modulo whitespace. In this way it is ensured that
(non)termination of the right TRS is proven.

\section{Results and Future Work}
\label{sec:experiments}
We ran extensive tests on the 1391 TRSs from version 5.0 of the termination
problems data base to evaluate the usefulness of \CeTA. As termination tool we
used \TTTT{}~\cite{TTT2}.

All tests have been performed on a server equipped with
eight dual-core AMD Opteron\textregistered{} 885 processors running
at a CPU rate of 2.6~GHz on 64~GB of system memory and with a time limit of
60~seconds for each TRS. 
The results can be seen in the following table
(where times are given in seconds and (proof-)sizes in kilobytes). 
\begin{center}
\begin{tabular}{l|rr@{~}l|rr@{~}l||rr@{~}lr@{~}l}
&\multicolumn{3}{c|}{\YES}
&\multicolumn{3}{c||}{\NO}
&\multicolumn{5}{c}{\CeTA{}}\\
&\multicolumn{1}{c}{\#}
&\multicolumn{2}{c|}{time (avg.)}
&\multicolumn{1}{c}{\#}
&\multicolumn{2}{c||}{time (avg.)}
&\multicolumn{1}{c}{\#}
&\multicolumn{2}{c}{time (avg.)}
&\multicolumn{2}{c}{size (avg.)}
\\\hline
& & & & & & \\[-2ex]
\TTTT{} & 572 & 460 & (0.80) & 214 & 147 & (0.68)
  & 786 & 114 & (0.14) & 41,145 & (52.35)
\end{tabular}
\end{center}
The \YES{} columns of the table denote termination proofs, whereas
the \NO{} columns denote found loops. That the numbers for
\YES{} and \NO{} sum up to 786, shows that \CeTA{} could certify every
proof generated by \TTTT{}.

For the future we are eager to combine our attempts for a certification XML
format with other approaches (like the \emph{termination certificates grammar} of
\Rainbow) to obtain a standard that can then be used by all termination tools.
Further, since we are working with automatically generated code, it would be
possible to combine our implementation with extracted code from other
formalizations in order to obtain a more powerful certifier.

\bibliographystyle{abbrv}
\bibliography{references}
\end{document}